\documentclass[12pt]{iopart}

\usepackage{iopams}  
\usepackage{amssymb}
\usepackage{siunitx}
\usepackage{graphicx}
\usepackage{transparent}
\usepackage{textcomp}
\usepackage{hyperref}
\usepackage[T1]{fontenc}
\usepackage[justification=raggedright]{caption}
\usepackage{cite}

\usepackage[mathlines]{lineno}

\begin{document}
\sisetup{range-phrase=-}
\sisetup{range-units=single}
\sisetup{per-mode=symbol} 

\title[Locust Simulation Software]{Locust:  C++ software for simulation of RF detection}

\author{A.~Ashtari~Esfahani$^1$, S.~B\"oser$^2$, N.~Buzinsky$^3$, R.~Cervantes$^1$, C.~Claessens$^2$, L.~de~Viveiros$^4$, M.~Fertl$^{1,2}$, J.~A.~Formaggio$^3$, L.~Gladstone$^5$, M.~Guigue$^6$, K.~M.~Heeger$^7$, J.~Johnston$^3$,
A.~M.~Jones$^6$, K.~Kazkaz$^8$, B.~H.~LaRoque$^6$,
A.~Lindman$^2$, E.~Machado$^1$, B.~Monreal$^5$, E.~C.~Morrison$^6$, J.~A.~Nikkel$^7$,
E.~Novitski$^1$, N.~S.~Oblath$^6$, W.~Pettus$^1$, R.~G.~H.~Robertson$^1$, G.~Rybka$^1$, L.~Salda\~na$^7$, V.~Sibille$^3$, M.~Schram$^6$, P.~L.~Slocum$^7$, Y.-H.~Sun$^5$, J.~R.~Tedeschi$^6$, T.~Th\"ummler$^9$, B.~A.~VanDevender$^6$, M.~Wachtendonk$^1$, M.~Walter$^9$,
T.~E.~Weiss$^3$, T.~Wendler$^4$, E.~Zayas$^3$}
\vspace{5mm}

\address{$^1$Center for Experimental Nuclear Physics and Astrophysics and Department of Physics, University of Washington, Seattle, WA  98195, United States of America}
\address{$^2$Institut f\"ur Physik, Johannes Gutenberg-Universit\"at Mainz, Mainz, Germany}
\address{$^3$Laboratory for Nuclear Science, Massachusetts Institute of Technology, Cambridge, MA 02139, United States of America}
\address{$^4$Department of Physics, Pennsylvania State University, State College, PA 16802, United States of America}
\address{$^5$Department of Physics, Case Western Reserve University, Cleveland, OH 44106, United States of America}
\address{$^6$Pacific Northwest National Laboratory, Richland, WA 99354, United States of America}
\address{$^7$Wright Laboratory and Department of Physics, Yale University, New Haven, CT 06520, United States of America}
\address{$^8$Lawrence Livermore National Laboratory, Livermore, CA 94550, United States of America}
\address{$^9$Institut f\"ur Kernphysik, Karlsruher Institut f\"ur Technologie, Karlsruhe, Germany}

\date{\today -- v2.0}

\vspace{10pt}

\begin{abstract}
The Locust simulation package is a new C++ software tool developed to simulate the measurement of time-varying electromagnetic fields using RF detection techniques.  Modularity and flexibility allow for arbitrary input signals, while concurrently supporting tight integration with physics-based simulations as input.  External signals driven by the Kassiopeia particle tracking package are discussed, demonstrating conditional feedback between Locust and Kassiopeia during software execution.  An application of the simulation to the Project~8 experiment is described.  Locust is publicly available at \url{https://github.com/project8/locust_mc}.
\end{abstract}

%
\vspace{2pc}

\section{Introduction}

\begin{figure}[htb]
\centering
\includegraphics[width=0.6\textwidth]{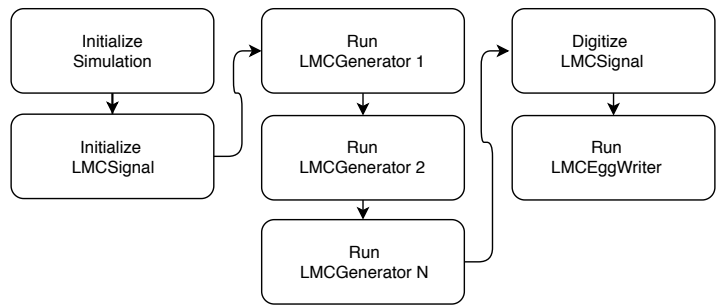}
\caption{\label{fig:flowchart}Flow chart of the Locust simulation software.  The modules in the center of the diagram, labeled LMCGenerator, are configurable to the user and function as stages in an RF receiver.  The LMCSignal object is first initialized, then passed through the LMCGenerator modules, and is finally digitized on completion of the simulation.   }
\end{figure}

The Locust software package~\cite{locust} is a simulation tool developed to model the response of an antenna and receiver to time-varying electromagnetic fields.  Its purpose is to generate data files formatted identically to those measured with an RF receiver and digitizer in the laboratory, thereby allowing for detailed calibration and simulation of physical measurements relying on RF detection techniques.  Written in C++, the software is modular and extensible to allow for algorithmic implementation of various RF receiver configurations.  The sensitive frequency range of Locust's detection is arbitrarily flexible.  The simulation accepts a calculated electromagnetic signal as input, the form of
which can be a sinusoidal waveform or an arbitrary externally-defined signal, and places it into a software object named ``LMCSignal''.  A collection of flexible classes called generators are available to the user for sequential configuration; these generators are named with the prefix ``LMCGenerator'' and represent components of an RF receiver.  Figure~\ref{fig:flowchart} shows a flow diagram with the classes that comprise the central function of Locust.

Locust presently has two main modes of operation available through appropriate selection of ``LMCGenerator 1'' in Figure~\ref{fig:flowchart}:  It can independently generate an ideal signal to use as input to its receiver, or it can instead wait periodically for an external electromagnetic signal from another software package.  Both modes of operation are represented in Figure~\ref{fig:flowchart}, with the functionality of the first generator block to be expanded to accommodate external signals in Section~\ref{sec:externalsignals}.
Section~\ref{sec:fakesignals} will discuss the first running mode using an internally generated signal.  Within the context of the Locust framework, the characteristics of the signal will be observed as it traverses the components of a simulated receiver, is digitized, and is then processed after Locust with the Katydid~\cite{katydid} analysis software. Section~\ref{sec:externalsignals} will refer to the second running mode in which Locust is integrated with another numerical model, an electron cyclotron orbit simulation performed with the Kassiopeia~\cite{Furse:2016fch} software package.

When considering how to model RF/microwave measurements, a question likely arises as to why a new simulation tool is needed.  The general reasons are for control and adaptability.  While there are presently several nonlinear time-domain RF circuit simulation tools that are commercially available (e.g. \cite{ads, mwo, ansyselec}), as well as EM field solvers (e.g. \cite{hfss,comsol,xfdtd,cst}), co-simulation frameworks that are driven by both tasks in cooperation are not as widely developed. Additionally many commercial simulation packages are at least partially closed-source, which limits their use as modular components in multistep calculations.  Further constraining the available options is the need for a straightforward, open-source interface allowing for arbitrary input from detailed physics models, as well as for data acquisition libraries that generate output with modifiable format.

The above criteria, along with a scarcity of suitable solutions, have motivated the development of Locust.  Locust supports highly adaptable computations in both the time and frequency domains, which allows for RF spectral interpretation of stochastic processes.  Of note is its initial application to the Project~8 experiment, an endeavor to constrain the effective electron neutrino mass by way of cyclotron radiation emission spectroscopy (CRES)~\cite{Asner:2014cwa, jphysg, pheno}.  

In the Project~8 experiment, decay electrons emit cyclotron radiation in a \SI{1}{\tesla} magnetic field.  Of particular interest is the region in the tritium beta spectrum near the \SI{18.6}{\keV} endpoint where the modification of the decay phase space by a non-zero effective neutrino mass becomes most significant.  Electrons emit radiation at frequency
\begin{equation}
f_\gamma = \frac{eB}{2\pi\gamma m_e},
\end{equation}
where $e$ is the charge of the electron, $B$ is the magnetic field strength, $m_e$ is the mass of the electron, the Lorentz factor is $\gamma=(1+K/m_e c^2)$, $K$ is the kinetic energy of the electron, and $c$ is the speed of light in vacuum.  As such, the experiment is sensitive to a range of electron energies that includes monoenergetic conversion electrons near \SI{30}{\keV} from $^{83\text{m}}$Kr.  As many of the first commissioning runs in Project~8 were tuned for detection of \SI{30}{\keV} electrons~\cite{jphysg}, Section~\ref{sec:externalsignals} shows a working Locust example comparing signals from \SI{30}{\keV} electrons in simulation and laboratory data.

\section{Receiver and digitizer simulation}\label{sec:fakesignals}
\subsection{Software flow}
The simplest type of input to Locust is an electric field sinusoidal in time.  The sinusoid can be defined as an arbitrary test signal by the user, or it can be configured to represent the properties of a theoretical model.  Non-sinusoids are also compatible with the simulation approach.  Referring to the diagram of classes in Figure~\ref{fig:flowchart}, the flow of the software begins in the top of the left column where the simulation properties are defined.  Properties include sampling frequency, record size, number of channels, and paths to output files.  Within these constraints the LMCSignal object is initialized as an array of complex voltages.  

Following initialization, the LMCSignal object is passed through generator blocks ``LMCGenerator 1'' through ``LMCGenerator N''.  Generators 1 through $N$ are ordered to allow initial population of the LMCSignal object in Generator 1, analogous to sampling a signal in the laboratory, followed by additional operations such as anti-alias filtering and downsampling in Generators 2 through $N$.  Certain generators, such as that for Gaussian noise as in Section~\ref{sec:noise}, can be configured without the need for filtering or downsampling and may be implemented either last in the sequence or alone.  In fact, Generators 1 through $N$ are not required by the software at all; if these blocks were removed from Figure~\ref{fig:flowchart}, the remaining simulation would output a time series of voltages all equal to 0.  Digitization and subsequent writing to disk occurs in the two blocks in the far right column.  The remainder of Section~\ref{sec:fakesignals} will describe the steps in the generation, filtering, digitization, and processing of a simulated laboratory signal in more detail.

\subsection{Simulation of signal generation}\label{sec:signal}
With a sinusoidal electric field as input, Locust first calculates the response of the antenna and stores the voltages in LMCSignal.  Voltage amplitude and phase are computed from the incident fields.  The voltage amplitude $V_0$ is derived from the incident electric field amplitude $|E_{inc}|$ by
\begin{equation}\label{eq:Vmag}
V_0 = |E_{inc}|/A_F
\end{equation}
where the antenna factor $A_F$, in units of 1/m, represents the gain of the antenna in converting from an incident electric field to an induced voltage across the antenna terminals.  If the input signal is known in units of power $P$ instead of V/m, then the conversion goes as
\begin{equation}\label{eq:VmagR}
V_0 = \sqrt{R}\sqrt{P}
\end{equation}
where $R$ is the antenna load impedance and is typically 50~$\Omega$.
The simulated voltage phase $\phi$ advances monotonically and discretely as
\begin{equation}\label{eq:phaseadv}
\Delta \phi(t) = 2 \pi f^\prime(t) \Delta t
\end{equation}
with $f^\prime$ as the time-dependent frequency calculated at the location of the antenna, and $\Delta t$ the time between voltage samples.  If there are $n>$1 propagating fields or noise fields incident at the antenna at time $t$, then the induced complex voltage $\tilde{V}_{RF}(t)$ is
\begin{equation}\label{eq:vsum}
\tilde{V}_{RF}(t) = V_{0,1}(t)e^{i\phi_1(t)} + V_{0,2}(t)e^{i\phi_2(t)} + \cdot\cdot\cdot + V_{0,n}(t)e^{i\phi_n(t)}
\end{equation}
where the $j$th propagating or noise field induces a voltage with magnitude $V_{0,j}(t)$ and phase $\phi_j(t)$.

The phase-sensitive voltages are represented by the in-phase $V_I$ and quadrature $V_Q$ components
\begin{eqnarray}
V_I(t)&=& V_0\cos(\phi(t)) \\
V_Q(t)&=& V_0\sin(\phi(t)).
\end{eqnarray}
While this definition of the voltages at the antenna terminals is useful for concrete discussion, in the simulation they are typically not calculated in the RF frequency band.  Instead, as will be explained in Section~\ref{sec:rx}, the voltages are first sampled in the intermediate frequency (IF) band in terms of their mixing product in LMCGenerator 1 in Figure~\ref{fig:flowchart}.  This improves efficiency by avoiding the need for sampling in the RF frequency band, which requires more intensive computing resources.

\subsubsection{Receiver}\label{sec:rx}

\begin{figure}[tb]
\centering
\includegraphics[width=0.8\textwidth]{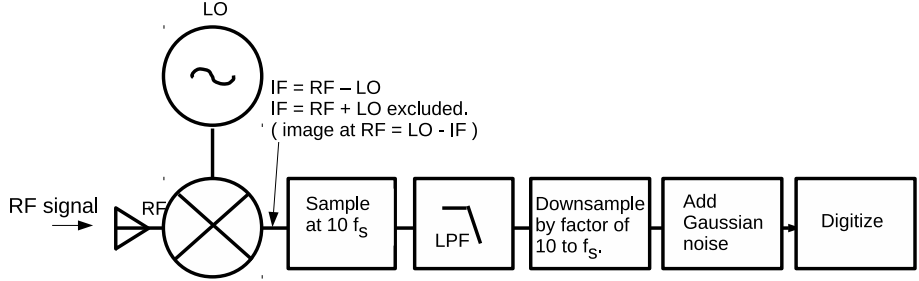}
\caption{\label{fig:rx}Block diagram of a receiver implemented algorithmically in Locust.  Each of the square blocks represents one generator as in Figure~\ref{fig:flowchart}.}
\end{figure}

A minimal receiver in Locust can in principle be a single generator in which voltages are sampled and then immediately digitized.  However, depending on experiment design and computing resource availability, more complexity may often be appropriate.  A realistic Locust receiver chain typically consists of a mixer with local oscillator at frequency $f_{LO}$, a low-pass filter, and a downsampling stage as shown in Figure~\ref{fig:rx}.  The downsampling reduces the sampling frequency by a factor of $M$ by discarding every $M$-1 out of $M$ samples (e.g.~\cite{oppenheim}).  The signal voltages are sampled as the mixing product
\begin{eqnarray}\label{eq:fullmixing}
V_I(t)&=& V_{RF}(t)\cos(\phi_{LO}(t)) \\
V_Q(t)&=& V_{RF}(t)\sin(\phi_{LO}(t)), \nonumber
\end{eqnarray}
where $V_{RF}(t)$ is the real part of the incident RF signal and $\phi_{LO}(t)$ is the phase of the local oscillator signal.  If $V_{RF}(t)$ is a sinusoidal waveform with phase $\phi_{RF}(t)$, then $V_{RF}(t)=V_0\cos(\phi_{RF}(t))$ and Equation~\ref{eq:fullmixing} can be written as
\begin{eqnarray}\label{eq:mixingsidebands}
V_I(t)&=& \frac{1}{2}V_0\left[\cos(\phi_{RF}(t)+\phi_{LO}(t)) + \cos(\phi_{RF}(t)-\phi_{LO}(t))\right] \\
V_Q(t)&=& \frac{1}{2}V_0\left[\sin(\phi_{RF}(t)+\phi_{LO}(t)) + \sin(\phi_{RF}(t)-\phi_{LO}(t))\right]. \nonumber
\end{eqnarray} 
The voltages in Equation~\ref{eq:mixingsidebands} can optionally be sampled without the upper sideband at phase $\phi_{RF}(t)+\phi_{LO}(t)$ as
\begin{eqnarray}\label{eq:sampling}
V_I(t)&=& \frac{1}{2}V_0\cos(\phi_{RF}(t)-\phi_{LO}(t)) \\
V_Q(t)&=& \frac{1}{2}V_0\sin(\phi_{RF}(t)-\phi_{LO}(t)). \nonumber
\end{eqnarray}
This omission of the upper sideband is analogous to attenuation by an ideal low-pass filter in the receiver chain.  Both $\phi_{RF}(t)$ and $\phi_{LO}(t)$ are typically calculated with Equation~\ref{eq:phaseadv}, using $f^\prime(t)$ and $f_{LO}$ to advance $\phi_{RF}(t)$ and $\phi_{LO}(t)$ in time, respectively.

Depending on experiment design, $\phi_{RF}(t)$ can accumulate at varying rates.  To accommodate any modulation, voltages are typically sampled at a rate 10$\times$ higher than the desired sampling frequency $f_S$.  This allows for accurate representation of unwanted high-frequency spurious signals between $f_S$/2 and 10$f_S$/2, making the signals suppressible using a low-pass filter with a threshold near the Nyquist frequency $f_S$/2~\cite{nyquist,shannon}.  Without the fast sampling rate the spurs can alias to frequencies below $\sim f_S$/2, where they will not be removed by the low-pass filter~\cite{nyquist,shannon}.  Following the low-pass filter is the downsampling stage, which reduces the fast sampling frequency back down to $f_S$.

A receiver chain such as the example above can be replicated in the simulation to accommodate multiple digitizer channels. The number of channels is specified in Locust by way of an externally-defined parameter.

\subsubsection{Gaussian noise}\label{sec:noise}
Complex random noise voltages are typically generated in the time domain and are added linearly to the existing signal voltages by way of Equation~\ref{eq:vsum}.  In the time domain the noise voltages are expressed as
\begin{eqnarray}
V_I(t) &=&  \sqrt{R}\sqrt{\text{0.5}} V_{I_{mag}}(t) \\
V_Q(t) &=&  \sqrt{R}\sqrt{\text{0.5}} V_{Q_{mag}}(t)
\end{eqnarray}
where $V_{I_{mag}}$ and $V_{Q_{mag}}$ follow a normal distribution with standard deviation $\sqrt{k_B T f_S}$ where $k_B$ is the Boltzman constant, $T$ is the noise temperature, and $f_S$ is the sampling rate after downsampling.

\subsubsection{Digitization and Signal Processing}\label{sec:daq}
\begin{figure}
\centering
\includegraphics[width=1.0\textwidth]{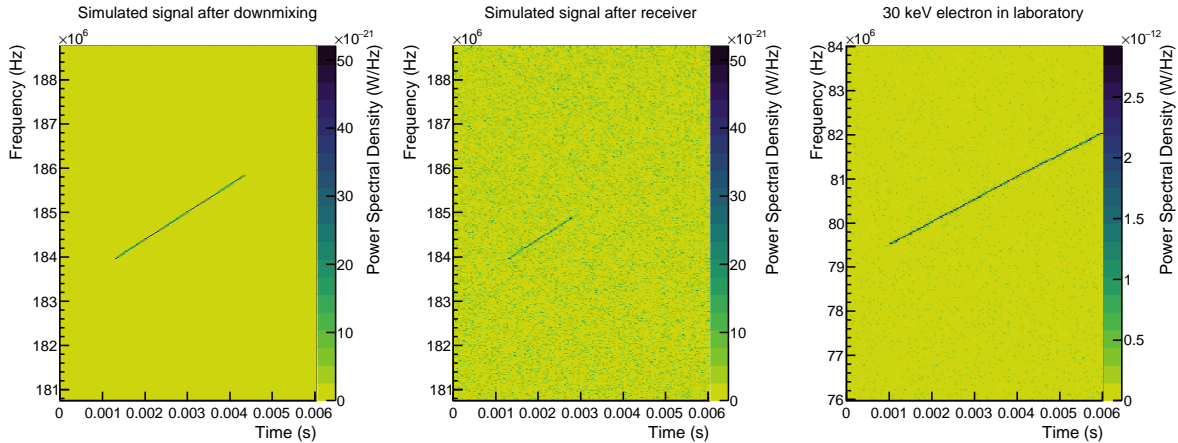}
\caption{\label{fig:faketrack} Spectrograms, processed with Katydid~\cite{katydid}, showing chirped tone signals generated either with Locust (left and middle panels) or measured in the Project~8 experiment~\cite{jphysg} (right panel).  The Locust spectra are simulated using two variations of the generators shown in Figure~\ref{fig:rx}.  The left panel shows an internal signal after downmixing and downsampling only, and the middle panel shows the same internal signal after downmixing, low-pass filtering, downsampling, and adding noise.  In the middle panel the upper edge of the low-pass filter is set to \SI{185}{\MHz}, above which the signal power is attenuated.  The right panel depicts a randomly chosen signal from a \SI{30.23}{\keV} electron measured in the first phase of the Project~8 experiment.  The scale of the power and frequency in the measured data differ from the simulation by a factor of receiver gain and by an offset in local oscillator frequency, respectively.}
\end{figure}

Before signals are written to disk they are digitized using parameters that can be matched to the data acquisition hardware being simulated (e.g. bit depth, and the voltage range and offset).  The I and Q components are digitized separately and stored as complex integer values.  The default file type is the ``Egg'' file, a file standard that was developed by the Project~8 collaboration based on the Hierarchical Data Format, version 5~\cite{hdf5}.  Egg files are designed for storing time-series data from one or more data sources such as digitizer channels.  They are created by Locust using the Monarch library~\cite{monarch}, a C++ library that includes an interface for writing and reading Egg files, as well as the documentation of the Egg file standard.  
Analysis of the digitized data is performed using the Katydid analysis software framework~\cite{katydid}.  From the data, Katydid can generate two-dimensional spectrograms of power in frequency and time that are useful for examination of Locust results.  Figure~\ref{fig:faketrack} shows an example of a spectrogram containing a Locust signal processed with Katydid using two different variations of the software receiver chain in Figure~\ref{fig:rx}, as well as a typical example of a signal measured in the Project~8 experiment.  Katydid is also used for more complex analyses to allow other direct comparisons between simulation and laboratory measurements.  An instance of this type of work will be shown below.

\subsection{Summary}
In this section the main purpose of Locust has been described as a simulation accurately representing a physical detector for RF signals.  Steps have been outlined in calculating voltages expected to be measured in response to an incident electromagnetic signal.  The above work provides a starting point for the discussion in the next section.
 
\section{Integrated radiation source simulation}\label{sec:externalsignals}
\begin{figure}[tb]
\centering
\includegraphics[width=0.3\textwidth]{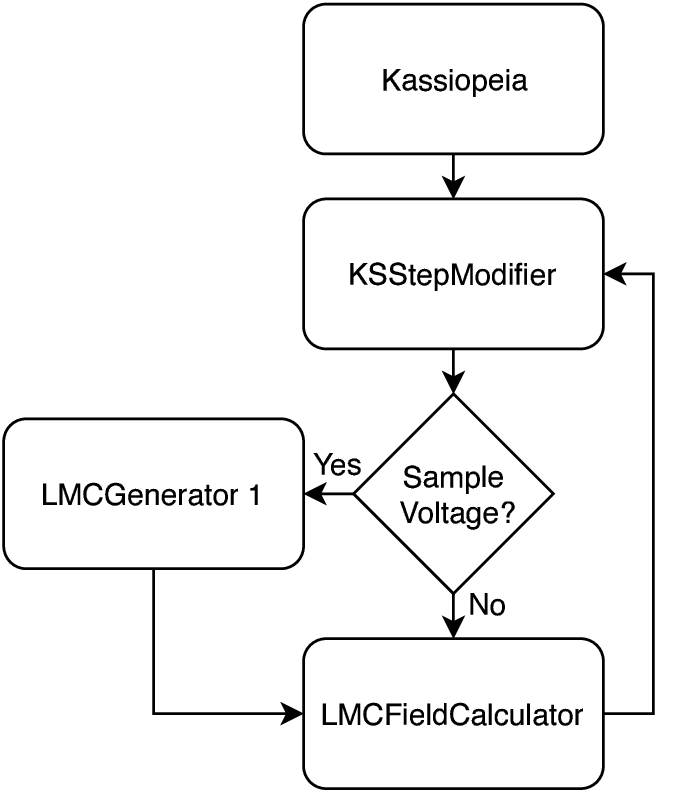}
\caption{\label{fig:kassflowchart}
Block diagram showing the flow of information between ``LMCGenerator 1'' in Figure~\ref{fig:flowchart} and the Kassiopeia software package.  The ``KSStepModifier'' class in Kassiopeia drives and accepts feedback from the Locust ``LMCFieldCalculator'' class, as described in the text.}
\end{figure}

Directly comparable to the running mode discussed above in which internal signals are defined and their detection is simulated, Locust can accept signals constructed with external software.  When running in this mode, a specialized generator serves as ``LMCGenerator 1'' in Figure~\ref{fig:flowchart}.  In the generator, the Locust C++ thread pauses and waits for information from the external software before populating each element of the LMCSignal time series as in Equation~\ref{eq:sampling}.  Next, just as in Section~\ref{sec:fakesignals}, the LMCSignal array is processed through an applicable receiver chain and is digitized.  To illustrate, Figure~\ref{fig:kassflowchart} shows a flow diagram in which the functionality of the block corresponding to ``LMCGenerator 1'' is expanded to interact with the external Kassiopeia software package~\cite{Furse:2016fch}.  The next section will describe this interface in more detail.

\subsection{Kassiopeia}\label{sec:kass}

One example of software that can be used to define external signals is the Kassiopeia simulation software package~\cite{Furse:2016fch}, compiled as a submodule within Locust.  First developed for the KATRIN experiment~\cite{katrin}, Kassiopeia is a software tool designed to compute electromagnetic fields in complex geometries, simultaneous with time-dependent solutions of particle states in position and momentum (${\bf x}$,${\bf p}$).  More specifically, it calculates particle trajectories stochastically in the presence of surface and gas interactions, while allowing for elastic and inelastic collisions.

Of the available particle interaction models in Kassiopeia, the focus of this work relies on the state (${\bf x}$,${\bf p}$) calculations of electrons with energies \SIrange{18}{30}{\keV} in a \SI{1}{\tesla} magnetic field.  This energy range is of interest to Project~8 because it is near the \SI{18.6}{\keV} endpoint of tritium beta decay, and near the \SI{30}{\keV} conversion electrons from $^{83\text{m}}$Kr.
The calculated electron trajectory ${\bf x}(t)$ is derived using the adiabatic approximation, which relies on conservation of the magnetic moment of the electron~(e.g. \cite{Furse:2016fch}).  The approximation is valid when variations in the electric and magnetic fields are minimal over each cyclotron orbit.  Kinematic constraints on the use of the approximation within Kassiopeia are discussed in~\cite{fursethesis}.  An eighth order Runge-Kutta integrator is used to solve the ordinary differential equation describing the trajectory.  Power $P$ lost by the electron to synchrotron radiation is then computed in Kassiopeia as~\cite{Furse:2016fch}
\begin{equation}\label{eq:kasspower}
P=\frac{\mu_0}{6\pi c}\frac{q^2}{m^2}(F^2_{||} + \gamma^2 F^2_\perp),
\end{equation}
where $F=dp/dt$ is the radiation reaction force, $F_{||}$ and $F_\perp$ are the components of $F$ parallel and perpendicular to the magnetic field,
\begin{equation}\label{eq:kasstrajectory}
\left.\frac{dp_\perp}{dt}\right|_{sync}=\frac{-\mu_0}{6\pi c}\frac{q^4}{m^3} \left|{\bf B}({\bf r}_c,t) \right|^2 p_\perp \gamma,
\end{equation}
$\mu_0$ is the permeability of free space, $q$ and $m$ are the charge and mass of the electron, and ${\bf B}({\bf r}_c,t)$ is the magnetic field at the guiding center of the motion ${\bf r}_c$.
In the adiabatic approximation $F_{||}$=0, implying that the cyclotron motion is responsible for the energy loss by the electron to radiation.  Together Equations~\ref{eq:kasspower} and \ref{eq:kasstrajectory} define the electron's energy loss and trajectory, computed numerically, as it moves through the calculated magnetic field.  The cyclotron frequencies reported for \SIrange{18}{30}{\keV} electrons in a \SI{1}{\tesla} field are near \SI{26}{\GHz}.
 
As Locust has been developed for the Project~8 collaboration, one of its initial applications has been to employ Kassiopeia's machinery to generate signals similar to those measured in the first phase of the Project~8 experiment~\cite{jphysg}.  The laboratory signals were measured at one end of a hermetic waveguide cell filled with gaseous $^{83\text{m}}$Kr, which emits conversion electrons at \SI{17.8243}{\keV}, \SI{30.4196}{\keV}, \SI{30.4723}{\keV}, and \SI{31.9370}{\keV}~\cite{nudat}, among others.
Emitted electrons were trapped magnetically in a \SI{0.9583}{\tesla} background magnetic field, and their cyclotron radiation was detected with an antenna and receiver.  

\begin{figure}
\centering
\includegraphics[width=0.49\textwidth]{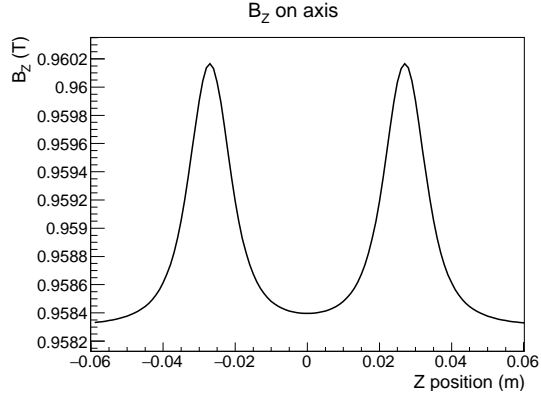}
\caption{\label{fig:fieldmap}Longitudinal component of the field map in the simulated magnetic trap, calculated with Kassiopeia~\cite{Furse:2016fch}.}
\end{figure}

In Locust and Kassiopeia the experiment is implemented with a magnetic trap having a longitudinal field map as in Figure~\ref{fig:fieldmap}, and a $^{83\text{m}}$Kr radioactive source contained within a rectangular WR42 (\SI{10.7}{\mm}$\times$\SI{4.3}{\mm}) waveguide cell of length \SI{10}{\cm}.  The Locust receiver is located at one end of the waveguide cell, and a reflecting waveguide short sits at the opposite end.  Electrons emitted with energies $>$\SI{30}{\keV} are tracked in time through the calculated magnetic field, while electrons with lower energies are terminated in the software.  This selection, in addition to a one degree wide restriction on pitch angle $\theta$ relative to the magnetic field, reduces overall computation time while allowing for reasonable agreement between measured data and simulation.  For comparison, pitch angles of electrons expected to be trapped with appreciable signal power in the context of CRES experiments are discussed in detail in~\cite{pheno}, and are consistent with the range of pitch angles selected in Locust.

\subsection{Waveguide modes}\label{sec:modecouplings}

The first step in the communication between Kassiopeia and Locust is to calculate how much energy is transferred from the Kassiopeia electron to the Locust receiver in the form of an external signal.  This computation happens iteratively between Kassiopeia and Locust, both at and between the times when voltages are sampled in Locust.  First, Kassiopeia reports the total power lost by the electron to cyclotron radiation in Equation~\ref{eq:kasspower}.  Next, Locust calculates the fraction of that total power deposited into the propagating mode or modes.  A mode that is able to propagate to the receiver is treated as an external signal; its detection in Locust occurs identically to that for the internally generated signals in Section~\ref{sec:fakesignals}.

The calculation proceeds as follows.
The time-averaged power $P^\pm$ radiated by the electron, given by Kassiopeia, is distributed into the sum of propagating and non-propagating waveguide modes~\cite{collin}.  Power propagating in both longitudinal directions is represented by ``$\pm$''.
In general notation similar to the discussion in~\cite{pozar}, the total time-averaged power is
\begin{equation}\label{eq:power}
P^\pm = \sum_\mu \frac{1}{Z_\mu}|A^\pm_\mu|^2,
\end{equation}
where $Z_\mu$ is the characteristic impedance of mode $\mu$ and $A^\pm_\mu$ is the time-averaged excitation amplitude of mode $\mu$ propagating into both longitudinal directions in the waveguide.  The fraction $\eta_\mu$ of the total power deposited into mode $\mu$ is then 
\begin{equation}\label{eq:powerfraction}
\eta_\mu = \frac{\frac{1}{Z_\mu} |A^{\pm}_\mu|^2}{P^\pm}. 
\end{equation}
The amplitudes of the propagating modes are derived from the Poynting theorem~\cite{jackson}
\begin{equation}\label{eq:poynting}
A_\mu = \int_V{{\bf J}\cdot{\bf E}_\mu d^3x},
\end{equation}
where ${\bf J}$ is the electron current, and $V$ is the volume enclosed by the waveguide walls and two surfaces enclosing the current distribution~\cite{jackson}. The transverse mode fields ${\bf E}_\mu$ are normalized at excitation as in \cite{jackson}
\begin{equation}\label{eq:norm}
\int_a{{\bf E}_\mu \cdot {\bf E}_\lambda} da = \delta_{\mu\lambda}
\end{equation}
where $\delta_{\mu\lambda}$ is the Kronecker delta function and $da$ is an element of the cross-sectional area of the waveguide.

While the time-averaged power is always distributed between propagating and nonpropagating modes, the instantaneous fraction of power deposited into the propagating modes is assumed to have a maximum near unity.  This approximation is applicable for this work given that amplitudes of non-propagating modes are decreased by propagation constants that are perturbed upward due to finite conductivity of the waveguide walls~\cite{collin}, and because non-propagating modes do not induce a direct response in the receiver electronics.  Indirectly they are present, as the energy deposited into non-propagating modes is contained implicitly in the calculation of trajectory and radiated power in Kassiopeia.  A statement similar to the latter is made in~\cite{collin} in that a change to the total power radiated into all waveguide modes by an electron should be accompanied by a change in the trajectory of the electron.

\begin{figure}
\includegraphics[width=1.0\textwidth]{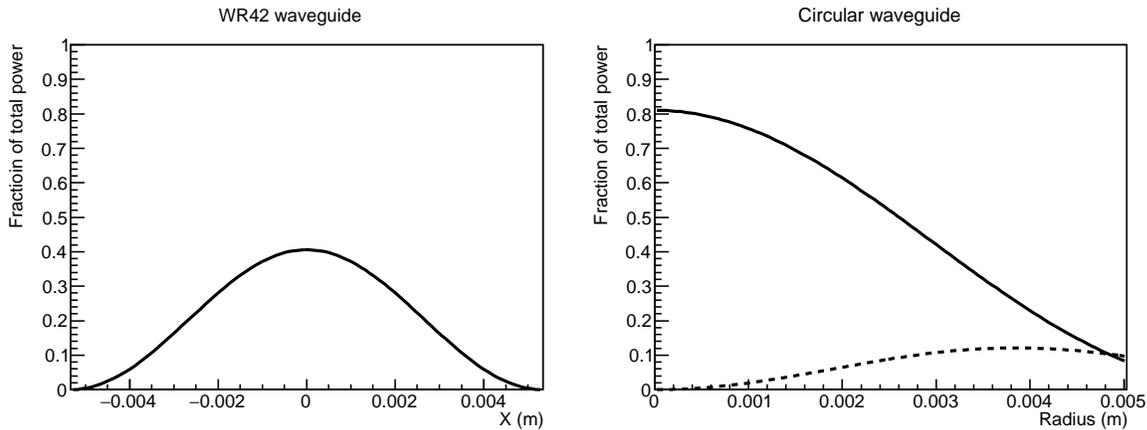}
\caption{\label{fig:modes}Power fractions contained in propagating modes at \SI{26}{\GHz} in the rectangular WR42 waveguide (left panel) and circular waveguide with diameter 0.396'' or \SI{10.0}{\mm} (right panel).  The rectangular case shows the TE10 mode across the long dimension of the waveguide cross section, and the circular case shows both the TE11 mode  (solid line) and the relatively low TM01 mode (dotted line) along the radius.}  
\end{figure}

As relevant to the first two phases of the Project~8 experiment, two waveguide geometries have presently been implemented in Locust.  Figure~\ref{fig:modes} shows the time-averaged fraction of power contained in the propagating modes for a \SI{10.7}{\mm}$\times$\SI{4.3}{\mm} rectangular WR42 waveguide cell and \SI{5.0}{\mm} radius circular waveguide cell at \SI{26}{\GHz}.  The circular waveguide in the right panel of Figure~\ref{fig:modes} supports two propagating modes, of which the TM01 power fraction is suppressed according to its wavenumber relative to that in the TE11 mode~\cite{collin}. Each of the calculations in Figure~\ref{fig:modes} are presently implemented in Locust, selectable by a parameter, and are referenced while the simulation is running.

\subsubsection{Mode propagation}\label{sec:modepropagation}

After the mode excitation, the propagating modes carry the appropriate fraction $\eta_\mu$ of the source energy in both longitudinal directions~\cite{jackson}.  Mode fields that propagate to the Locust receiver are processed as an external signal.  Any mode field that does not propagate to the receiver does not induce a voltage.  Finally, a mode $\mu$ field that propagates away from the electron, and then reflects back to the location of the electron, sums with the ongoing field excitation by the electron.  The outcome of the latter case, described in the next section, has been uniquely suited within the Locust framework to explain a subset of measurements in the Project~8 experiment~\cite{Asner:2014cwa}.

\subsubsection{Waveguide back-reaction} \label{sec:modeconfiguration}

In the simulation, the mode configuration in the waveguide at the location of the electron affects the amount of power radiated in Equation~\ref{eq:power}.  This back-reaction by the field on the electron behaves as stimulated emission (e.g. \cite{townes, Twiss:1958cd, hirshfield, fain}) that is self-induced by the electron in the waveguide.  A similar back force between a radiating electron and its induced mode fields has been observed in resonant cavities~\cite{purcell, gabrielse, gabrielse2}; this type of effect has also been widely employed in lasers with resonant cavities.  In~\cite{gabrielse} the back force is described quantitatively in terms of image charges that constrain the induced cavity mode fields.  And, in~\cite{gabrielse2} spontaneous emission by a radiating electron is observed to be inhibited by self-induced fields in a resonant cavity.

In the Locust simulation the effect due to the waveguide back-reaction is computed by way of time-dependent energy conservation in Equations~\ref{eq:poynting} and~\ref{eq:norm}.  As time advances in the simulation, energy is exchanged iteratively between the electron and the waveguide mode field that it induced previously.  First, in Equation~\ref{eq:norm} the mode fields are normalized at their initial excitation, but are not normalized again if they return to the electron by way of a reflection.  The reflected field is instead added everywhere to the existing normalized mode ${\bf E}_\mu$, giving ${\bf E}^\prime_\mu$.  The time-dependent sum at the electron is
\begin{equation}\label{eq:eprime}
    |{\bf E}^\prime_\mu(t)| = \cos(0) + \cos(\xi(t)),
\end{equation}
where $\cos(0)$ represents the normalized field induced at the electron and $\cos(\xi(t))$ is the reflected field with phase $\xi(t)$.  Neglecting propagation times for small geometries, $\xi(t)$ advances from zero at the electron's position to $\xi(t)>0$ according to propagation distance and wavelength.  If the reflector is a conductor, as in the calculation to be discussed in Section~\ref{sec:example}, $\xi(t)$ is calculated as
\begin{equation}\label{eq:xi}
\xi(t) = \pi/2 + 2\pi (|z(t)| + D)/\lambda^\prime(t).
\end{equation}
Otherwise it is typically
\begin{equation}\label{eq:xiTM}
\xi(t) = 2\pi\cdot2(D-|z(t)|)/\lambda^\prime(t).  
\end{equation}

In the expression for $\xi(t)$, $z(t)$ is the longitudinal position of the electron, $D$ is the distance from the center of the magnetic trap to the reflector, and $\lambda^\prime$ is the Doppler-shifted wavelength calculated at the reflector using the group velocity of the propagating mode.  Longitudinal symmetry is enforced with $|z(t)|$, as needed for agreement with the data, and is consistent with the indication that $A^+_\mu=A^-_\mu$ in~\cite{jackson,collin}.  In Locust, Equations~\ref{eq:eprime}-\ref{eq:xiTM} are calculated in the ``LMCFieldCalculator'' block shown in  Figure~\ref{fig:kassflowchart}.

With the reflected fields Equation~\ref{eq:poynting} becomes
\begin{equation}\label{eq:poyntingprime}
A_\mu^\prime = \int_V{{\bf J}\cdot{\bf E}^\prime_\mu d^3x} = |{\bf E}^\prime_\mu| A_\mu,
\end{equation}
which induces a correction to the total power $P^\pm$ radiated by the electron in Equation~\ref{eq:power}.  Specifically, $P^\pm_\mu$ is adjusted from its nominal value without reflections in Section~\ref{sec:modecouplings} to its new value
\begin{equation}
P^{\pm\prime}_\mu = \frac{1}{Z_\mu} |A^{\pm\prime}_\mu|^2.
\end{equation}
Adjusting the total power $P^\pm$ by replacing $P^{\pm}_\mu$ with $P^{\pm\prime}_\mu$,
\begin{eqnarray}\label{eq:powerreflections}
P^{\pm\prime} &=& P^\pm - \frac{1}{Z_\mu} |A^{\pm}_\mu|^2 + \frac{1}{Z_\mu} |A^{\pm\prime}_\mu|^2 \\
              &=& P^\pm\left(1 - \frac{\frac{1}{Z_\mu} |A^{\pm}_\mu|^2}{P^\pm} + \frac{\frac{1}{Z_\mu} |A^{\pm\prime}_\mu|^2}{P^\pm}\right),
\end{eqnarray}
where the last two terms each represent the fraction of power $\eta_\mu$ radiated into mode $\mu$ as in Equation~\ref{eq:powerfraction}.  In Locust Equation~\ref{eq:powerreflections} is applied in the loop spanning the ``KSStepModifier'' and ``LMCFieldCalculator'' classes shown in Figure~\ref{fig:kassflowchart}.  

Equation~\ref{eq:powerreflections} probably appears unstable in that the power $P^\pm$ depends on its own history.  Likewise
in~\cite{gabrielse} the problematic outcome of an infinity from the calculated interaction of an electron with its self-energy in a resonant cavity is identified, and is subsequently renormalized by carefully subtracting the field contribution from selected image charges.  While the experiment detailed in~\cite{gabrielse} differs from that described here, the computational difficulties arising from the back force can be analogous. Fortunately, in Locust the relevant mode fields in Equation~\ref{eq:powerreflections} are separated in time, which makes their effect finite when calculated iteratively.

Another implication of Equation~\ref{eq:powerreflections} is that the electron radiates more (less) than it does in free space if $\left|{\bf E}_\mu^\prime(t)\right|/\left|{\bf E}_\mu(t)\right|$ is more (less) than unity.  Thus the power radiated by the electron, initially reported in Kassiopeia and represented in Equation~\ref{eq:power}, is altered at each trajectory step according to the time-dependent sum at the electron's location of both the normalized and the reflected propagating mode fields.  The resulting trajectory of the electron computed in Kassiopeia changes in response to this adjusted energy loss.  As a related consequence, the amplitude of the propagating mode fields also scale iteratively with the total radiated power as in Section~\ref{sec:modecouplings}.  Each of the two critical quantities (radiated power and propagating power) are calculated repeatedly while the simulation is running; the electron's energy loss is adjusted according to the mode configuration after every tracking step in Kassiopeia, while the propagating mode amplitude is calculated only at the relatively infrequent times when voltages are sampled at the receiver.

\section{Results}\label{sec:example}

Simulated and experimental results are compared using identical analysis chains with the Katydid~\cite{katydid} software.  Signals from electron ``tracks'' of power in frequency and time, similar to those in the spectrograms of Figure~\ref{fig:faketrack}, are identified and their characteristics recorded.  The sampling rate $f_S$ in each data set is \SI{200}{\MHz}.  The receivers are both tuned to measure the \SI{30.420}{\keV} and \SI{30.472}{\keV} conversion electrons from a $^{83\text{m}}$Kr source.  In the simulation the local oscillator frequency is set to \SI{25.3106}{\GHz}.  The simulated magnetic field map is shown in Figure~\ref{fig:fieldmap} and is consistent with that used in the experiment. 
 
The intermediate steps in the analysis are:
\begin{enumerate}
\item Perform a Fast Fourier Transform of each 8192-sample-long time series;
\item To compensate for frequency-dependent variations in gain, integrate the spectrogram over the full acquisition in time to obtain the average power spectrum and approximate it with a spline fit;
\item Impose a cut on signal-to-noise ratio by selecting high-power bins above a threshold set by spline fit to the background;
\item High-power bins with too few neighboring points are removed and the surviving bins are clustered together using the Density-Based Spatial Clustering of Applications with Noise ({\small DBSCAN}) algorithm~\cite{dbscan};
\item Clusters of points are analyzed for linear structure using a Hough Transform~\cite{hough}, resulting in track objects.
\end{enumerate}

\begin{figure}
\centering
\includegraphics[width=0.49\textwidth]{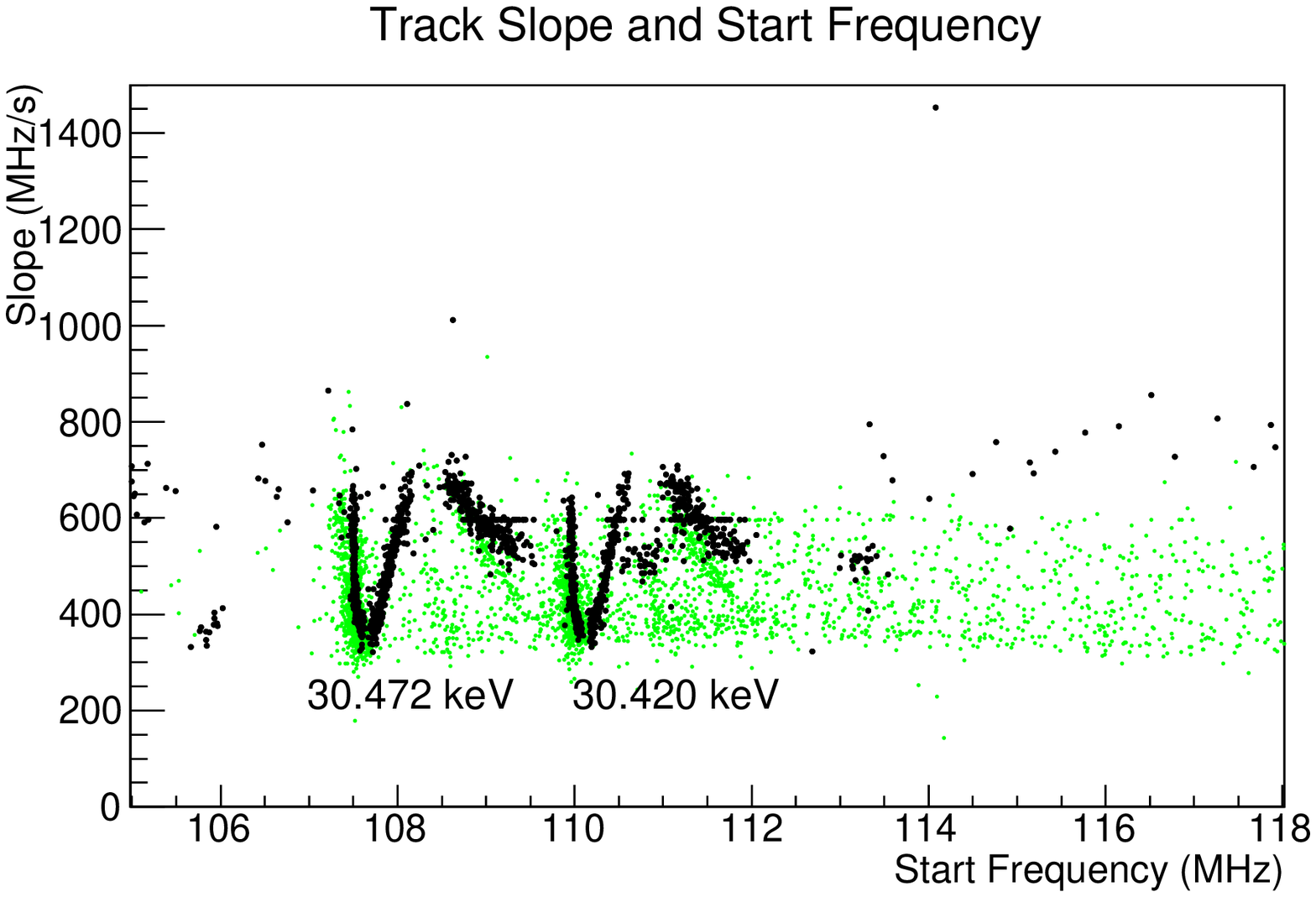}\hspace{-0.2in}{\small(a)}
\includegraphics[width=0.49\textwidth]{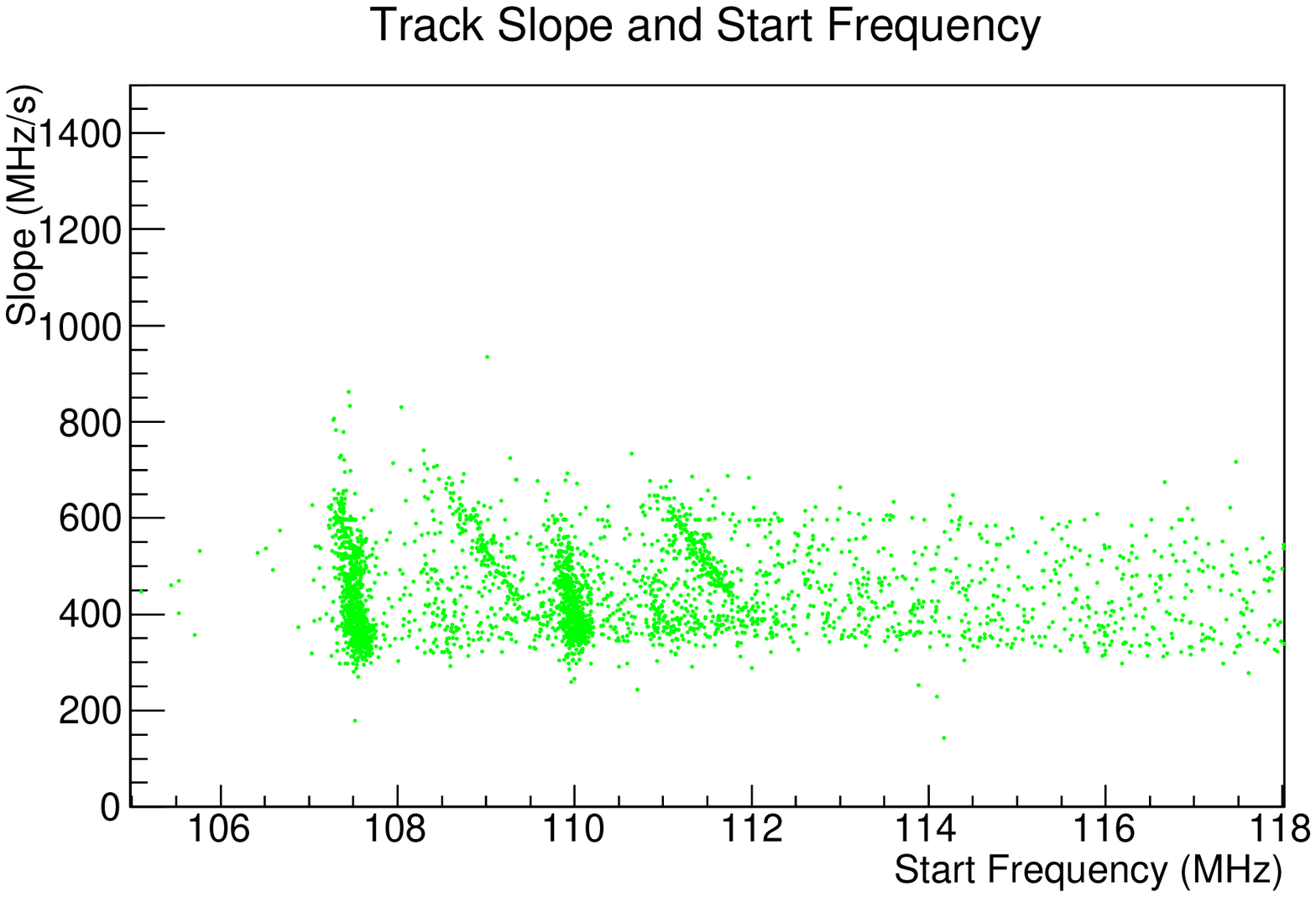}\hspace{-0.2in}{\small(b)}
\includegraphics[width=0.49\textwidth]{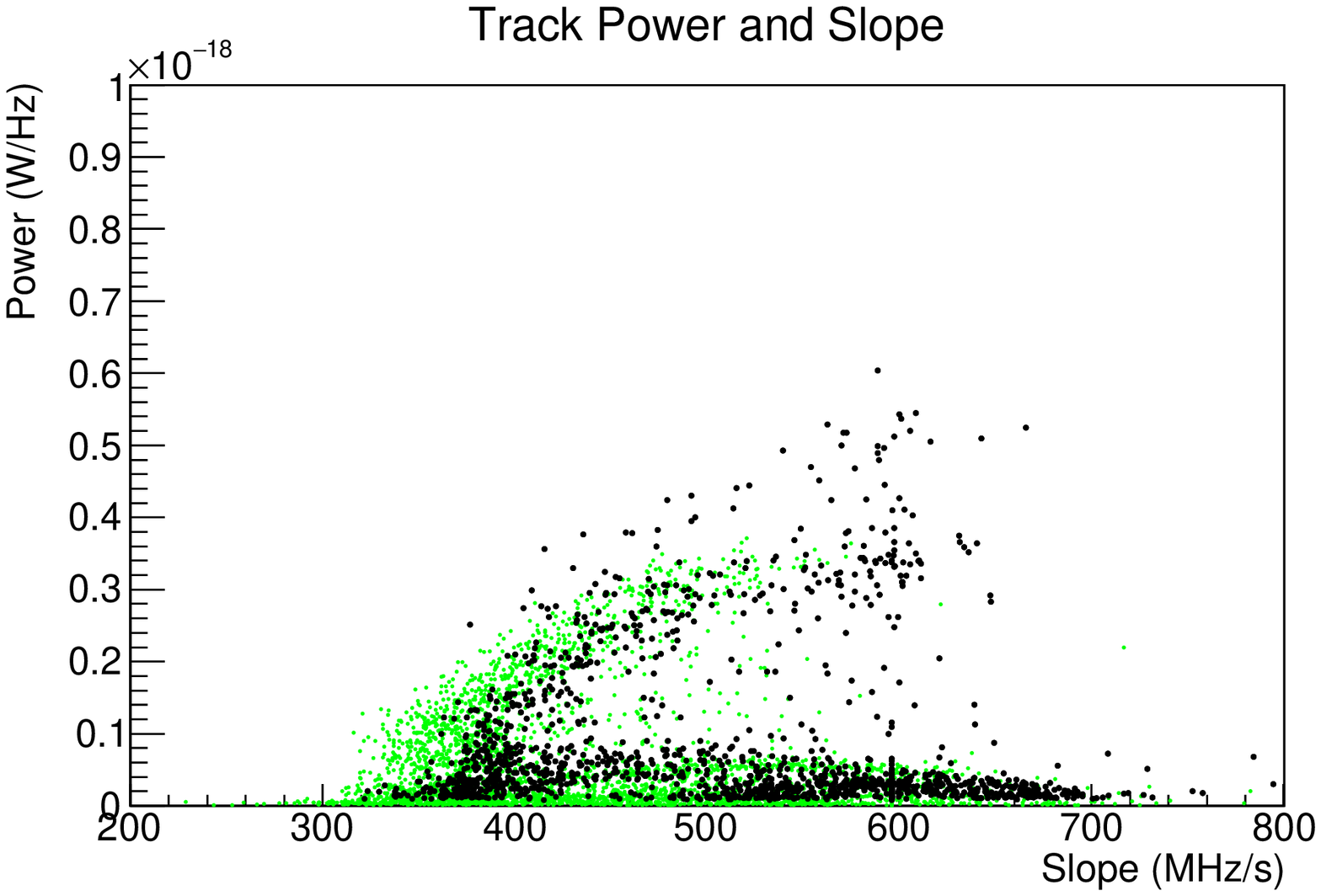}\hspace{-0.2in}{\small(c)}
\includegraphics[width=0.49\textwidth]{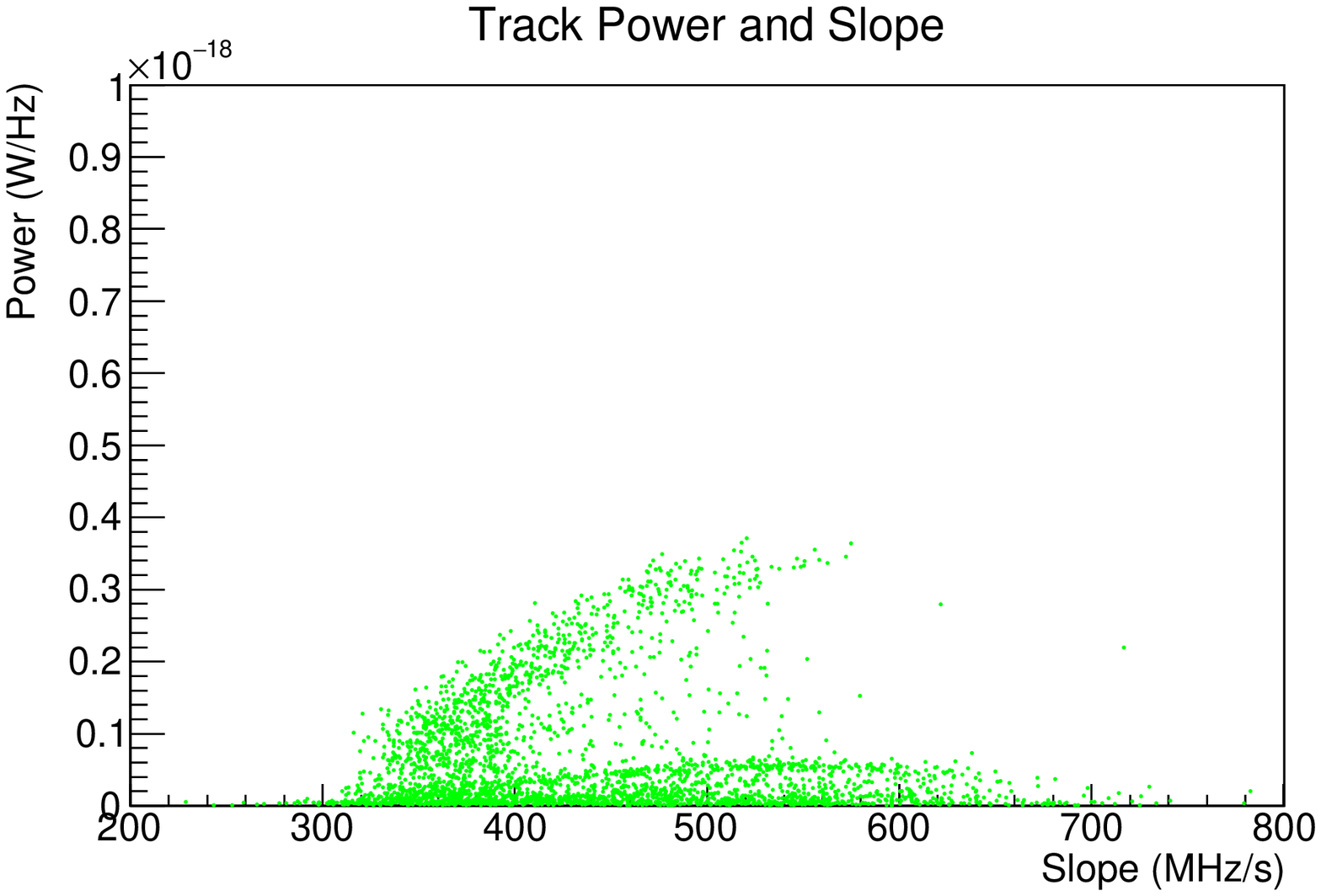}\hspace{-0.2in}{\small(d)}
\caption{\label{fig:Phase1Sim}  Comparison between data and simulation for \SI{30.420}{\keV} and \SI{30.472}{\keV} conversion electrons emitted from a $^{83\text{m}}$Kr source.  Electron track characteristics extracted from simulation (black) and measurement (green) are plotted.  The top two panels (a) and (b) show track slope plotted against track starting frequency for the data and simulation.  The lower two panels (c) and (d) show the track power, with that in the data scaled downward by a factor of the inferred laboratory receiver gain, plotted against the track slope.  Receiver gain is extracted at 93~dB, but is probably lower due to differences in track duration between data and simulation that are driven by computation time.  Resolution bandwidth is 24 kHz.}
\end{figure}

Characteristics of the analyzed electron tracks are shown in Figure~\ref{fig:Phase1Sim}.  The track slope corresponds to the track's rate of change of frequency in time, and the track power is derived from the summed amplitudes in the track as plotted in the spectrogram.  The starting frequency of each track is derived from the frequency of the first bin in the track.  The agreement between data and simulation is reasonable.  There is an apparent excess of points with slope near \SI{600}{\MHz\per\second} in both data and simulation.  This behavior is a binning effect in the 2D spectrograms in frequency and time that is more pronounced for events with shorter duration.  Small deviations of the simulation from the data are expected due to substantial sensitivity to the mode configuration in Section~\ref{sec:modeconfiguration} as well as to the magnetic field map.  Accordingly, changing the longitudinal position of the reflector by \SI{0.5}{\mm} is enough to detune the agreement noticeably by eye, as is altering the magnetic trap coil currents by \SI{0.02}{\ampere}.  The noise temperature applied in Figure~\ref{fig:Phase1Sim}, inferred in the simulation from the data, is \SI{15}{\kelvin}.  

Although there is qualitative agreement in Figure~\ref{fig:Phase1Sim}, the simulation result deviates from the laboratory data in several places.  The differences are more pronounced in regions where the track power is low.  For example, the upstroke of the black ``V''-shaped feature in the simulated points in panel (a) is absent from the measured data.  Additionally, the tilted band in panel (c) appears less intense in the simulation than it does in the measured data, and the range of slopes in the data in panel (c) extends slightly lower than that in the simulation.

Given the extent of the differences between data and simulation, some discussion is needed on what kind of additional work could improve the agreement.
Because the magnetic field and trap dimensions have already been carefully tuned to simulate this result, effort toward improving the agreement should begin with several other concrete steps.  First, the restriction on pitch angles in the simulation, presently applied with arbitrary uniformity from 89\si{\degree} to 90\si{\degree}, can be adjusted after straightforward development in software and computing to allow numerically-calculated intensities that vary across the same range.  Second, propagation times that are presently neglected in Equations~\ref{eq:xi} and \ref{eq:xiTM} could be built into the calculation.  Third, after implementing the latter improvements, higher statistics in the simulation would likely increase the broadband noise points in the simulated result as is depicted in the measured spectra in panels (a) and (b).  Finally, the range of slopes in the simulation might be extensible to lower values through a closer examination of the power correction in Equation~\ref{eq:powerreflections}.  Presently it is derived instantaneously from time-averaged amplitudes.  If possible, it would be worthwhile to check whether the range of simulated slopes could be extended lower with a more precise representation of the time dependence in the mode configuration.

While the present simulation is probably too discrepant from the data to allow detailed quantitative analysis, it provides important information on modeling time-dependent behaviors in CRES experiments.  In particular, as stated above, the interaction between the electron and its reflected mode field in Section~\ref{sec:modeconfiguration} is treated here with approximate time dependence.  If further stochastic particle-driven computations are of interest, a reasonable goal would be refinement of the temporal interpretation of this and other related effects.  At the same time, the simulation can be used to evaluate the impact of systematic effects on the experiment, and to optimize the granularity of time-dependent calculations that are needed to model it.

\section{Conclusion}
The Locust software package simulates the detection of RF signals comparable to signals measured in the laboratory.  Its range of applications extends from user-defined standalone test signals to field excitations computed from simulated particles.  It is useful both as a framework for investigating the feasibility of experiments in the laboratory, and as a tool for offline data analysis.  Additional work with Locust is underway to study and optimize future, larger-scale experiments in support of the Project~8 collaboration.   The source code along with examples and instructions for installation is available for download from \url{https://github.com/project8/locust_mc}.

\section{Acknowledgments}
This material is based upon work supported by the following sources: the U.S. Department of Energy Office of Science, Office of Nuclear Physics, under Award No.~DE-SC0011091 to the Massachusetts Institute of Technology (MIT), under Early Career Award No. DE-SC0019088 to Pennsylvania State University, under the Early Career Research Program to Pacific Northwest National Laboratory (PNNL), a multiprogram national laboratory operated by Battelle for the U.S. Department of Energy under Contract No.~DE-AC05-76RL01830, under Award No.~DE-FG02-97ER41020 to the University of Washington (UW), and under Award No.~DE-SC0012654 to Yale University; the National Science Foundation under Award Nos.~1205100 and 1505678 to MIT; Laboratory Directed Research and Development (LDRD) at Lawrence Livermore National Laboratory (LLNL) 18-ERD-028, LLNL-JRNL-789200, prepared by LLNL under Contract DE-AC52-07NA27344; the MIT Wade Fellowship; the LDRD Program at PNNL; and the UW Royalty Research Foundation. A portion of the research was performed using Research Computing at PNNL. The isotope used in this research was supplied by the United States Department of Energy Office of Science by the Isotope Program in the Office of Nuclear Physics. We further acknowledge support from Yale University, the PRISMA Cluster of Excellence at the University of Mainz, and the Karlsruhe Institute of Technology (KIT) Center Elementary Particle and Astroparticle Physics (KCETA).
\section{References}
\bibliographystyle{iopart-num}
\bibliography{LocustBib}

\end{document}